\documentclass[]{iopart}
\usepackage{iopams}
\usepackage{setstack}
\usepackage[]{graphicx}
\usepackage{color}

\usepackage{mathbbol}
\usepackage{amssymb}             
\DeclareSymbolFontAlphabet{\amsmathbb}{AMSb}%

\newcommand{\colr}{}

\newcommand{\tu}{\tilde{u}}
\newcommand{\Real}{{\rm{Re}}}
\newcommand{\Imag}{{\rm{Im}}}

\newcommand{\bp}{{\bi{p}}}
\newcommand{\brr}{{\bi{r}}}

\begin{document}

\title[]{\colr Wavefront errors in a two-beam interferometer}
\author{G Mana, E Massa and C P Sasso}
\address{INRIM - Istituto Nazionale di Ricerca Metrologica, Str.\ delle Cacce 91, 10135 Torino, Italy}
\ead{g.mana@inrim.it}

\begin{abstract}
The paper deals with the impact of wavefront errors, due to the optical aberrations of a two-beam interferometer, on the period of the travelling fringe observed by integrating the interference pattern. A Monte Carlo simulation of the interferometer operation showed that the fringe-period estimate is unbiased if evaluated on the basis of the angular spectrum of the beam entering the interferometer, but the wavefront errors increase the uncertainty.
\end{abstract}

\submitto{Metrologia}


\section{Introduction}
In length metrology by optical interferometry, the wavefront errors affect the period of the interference signal. The calibration of lasers against frequency standards achieves relative uncertainties smaller than $10^{-10}$, but it is not possible to trace back the wavelength to the frequency via the plane-wave dispersion equation. The relevant corrections have been extensively investigated in the literature \cite{Dorenwendt:1976,Mana:1989,Bergamin:1994,Cordiali:1997,Niebauer:2003,Cavagnero:2006,Robertsson:2007,Fujimoto:2007,Dagostino:2011,Andreas:2011,Andreas:2012,Andreas:2015,Andreas:2016,Mana:2018}. When the interfering wavefronts differ only by the propagation distances through the interferometer arms, the fractional wavelength difference -- which, typically, ranges from parts in $10^{-7}$ to parts in $10^{-9}$ -- is proportional to the square of the beam divergence which, for arbitrary paraxial beams, is proportional to the trace of the second central-moment of the angular power-spectrum \cite{Bergamin:1999,Mana:2017a}.

Characterizations of the laser beams leaving a combined x-ray and optical interferometer brought into light wavefront and wavelength ripples having a spatial bandwidth of a few mm$^{-1}$ and amplitudes as large as $\pm 20$ nm \cite{Balsamo:2003} and $\pm 10^{-8}\lambda$ \cite{Sasso:2016}, respectively, which might have a detrimental effect on the accuracy of the measurements. Since the differential wavefront-errors -- i.e., a non-uniform phase profile of the interference pattern -- cannot be explained by aberrations of beam feeding the interferometer, we carried out an analysis of the effect of wavefront aberrations due to the interferometer optics.

In section \ref{theory}, we outline the mathematical framework needed to model two-beam interferometry and paraxial propagation and show a {\colr one-dimensional} analytical calculation of the difference of the fringe period from the plane-wave wavelength. Eventually, we report about a Monte Carlo {\colr two-dimensional} calculation of the fringe period in the presence of wavefront errors caused by the interferometer optics.

\section{Mathematical model}\label{theory}
\subsection{Phase of the interference signal}
The interferometer slides two beams, $u_0(\brr;z+s)\exp(-\rmi kz)$ and $u_1(\brr;z)\exp(-\rmi kz)$, one with respect to the other by a distance $s$ while keeping them coaxial. By leaving out the $\exp(-\rmi kz)$ term of the optical fields -- where $k=2\pi/\lambda$ is the plane-wave wave number and $z$ the propagation distance -- and assuming and infinite detector, the integrated interference signal is
\begin{eqnarray}
 S(s) &= &\int_{-\infty}^{+\infty} \int_{-\infty}^{+\infty} |u_0(\brr;s)+u_1(\brr;0)|^2 \, \rmd \brr \nonumber \\
      &= & \int_{-\infty}^{+\infty} \int_{-\infty}^{+\infty} |\tu_0(\bp;s)+\tu_1(\bp;0)|^2 \, \rmd \bp ,
\end{eqnarray}
where we reset the origin of the $z$ axis, $\brr$ is a position vector in the detector plane (orthogonal to the $z$ axis), $\tu_0(\bp;s)$ and $\tu_1(\bp;0)$ are the angular spectra of the interfering beams \cite{Goodman:1996}, and $\bp$ is the wavevector of the angular spectra basis, $\exp(-\rmi \bp \bi{r})$.

The phase of the integrated interference pattern in excess (or defect) with respect to $-ks$ is \cite{Bergamin:1999}
\begin{equation}\label{phase}
 \Phi(s) = \arg\big[\Xi(s)\big] ,
\end{equation}
where
\begin{equation}\label{Xi}
 \Xi(s) = \int_{-\infty}^{+\infty} \tu_1^*(\bp;0) U(\bp;s) \tu_0(\bp;0)\, \rmd \bp
\end{equation}
is the interference term of the integrated intensity. In (\ref{Xi}), we used $\tu_0(\bp;s) = U(\bp;s) \tu_0(\bp;0)$, where
\begin{equation}
 U(\bp;s)= \exp\left(\frac{\rmi p^2 s}{2k}\right) ,
\end{equation}
is the reciprocal space representations of the paraxial approximation of the free-space propagator and $p^2=|\bp|^2$. The fringe period is
\begin{equation}\label{error}
 \lambda_e  = \lambda \bigg(1 + \frac{1}{k} \frac{\rmd \Phi}{\rmd s} \bigg|_{s=0} \bigg) ,
\end{equation}
where the sign of the derivative is dictated by the negative sign chosen for the plane-wave propagation. {\colr It must be noted that, since $U^*(\bp;z)U(\bp;z+s) = U(\bp;s)$, the interfering beams can be propagated by the same distance $z$ without changing (\ref{Xi}) and, consequently, $\lambda_e$. Therefore, (\ref{error}) depends only on the length difference of the interferometer arms, not on the detection-plane distance from, for instance, the beam waist.}

The interferometer recombines the light beams after delivering it through arms of different optical lengths. We consider the case when the interferometer arms have the same length; an analysis of the fringe phase and period as a function of the arm difference is given in \cite{Cavagnero:2006}. However, we want to allow the interferometer arms to deviate from perfection. Therefore, $\tu_1(\bp;0)$ and $\tu_0(\bp;0)$ are intrinsically different, meaning that they cannot be made equal by freely propagating one of the two, and, as implicitly assumed in (\ref{Xi}), the aberrations occur after the beam splitting but before the interferometer mirrors.

\subsection{Propagation of the wavefront errors}
To give an analytical one-dimensional example, let the complex amplitudes of the direct space representation of the interfering beams differ by a small wavefront error $\varphi(x)$, that is,
\begin{equation}\label{u-series}
 u_1(x) = u_0(x) \rme^{\rmi \varphi(x)} \approx u_0(x) \big[ 1 + \rmi\varphi(x) - \varphi^2(x)/2 \big] ,
\end{equation}
where we omitted the $z=0$ specification, and let
\begin{equation}\label{gauss}
 u_0(x) = \left(\frac{2}{\pi w_0^2}\right)^{1/4} \rme^{-x^2/w_0^2}
\end{equation}
be a normalized Gaussian beam. Since we are interested to small sliding distance with respect to the Rayleigh length -- that is, $ks\theta^2/2 \ll 1$, where $\theta$ is the beam divergence -- it is convenient to use a finite difference approximation of the $z$ derivative in the paraxial wave equation and the first-order approximation,
\begin{equation}
 U(x;s) \approx 1 - \frac{\rmi s \partial^2_x}{2 k} ,
\end{equation}
of the direct-space representation of the free-space propagator. Hence,
\begin{equation}\label{Xix}
 \Xi(s) = \int_{-\infty}^{+\infty} u_1^*(x) \left( 1 - \frac{\rmi s \partial^2_x}{2 k} \right) u_0(x)\, \rmd x .
\end{equation}
It is worth noting that, since the $-\partial^2_x$ operator is self-adjoint, it does not matter what of the interfering beams is slid. Therefore, for the convenience of the $\Xi(s)$'s computation, we choose to propagate $u_0(x)$. By using (\ref{u-series}) and carrying out the integrations in (\ref{Xix}), we obtain \cite{Mathematica}
\begin{numparts}\begin{equation}\fl
 \Real[\Xi(s)] = \sqrt{\pi} - \frac{s}{kw_0^2} \int_{-\infty}^{+\infty} \rme^{-\xi^2}(1-\xi^2)\varphi(\xi)\, \rmd \xi -
 \frac{1}{2} \int_{-\infty}^{+\infty} \rme^{-\xi^2}\varphi^2(\xi)\, \rmd \xi
\end{equation}
and
\begin{equation}\fl
 \Imag[\Xi(s)] = \frac{\sqrt{\pi} s}{2kw_0^2} - \frac{s}{2kw_0^2} \int_{-\infty}^{+\infty} \rme^{-\xi^2}(1-\xi^2)\varphi^2(\xi)\, \rmd \xi
 + \int_{-\infty}^{+\infty} \rme^{-\xi^2}\varphi(\xi)\, \rmd \xi ,
\end{equation}\end{numparts}
where $\xi=\sqrt{2}x/w_0$.

\subsection{Fractional error}\label{analytical}
The plane-wave wavelength is shorter than the fringe period $\lambda_e$ as defined in (\ref{error}); the fractional difference is \cite{Mathematica}
\begin{eqnarray}\label{corre-1}\fl\nonumber
 \frac{\Delta \lambda}{\lambda} \approx &\frac{\theta^2}{8} \bigg[ 1 +
 \frac{1}{\pi} \int_{-\infty}^{+\infty} \rme^{-\xi^2}(1-2\xi^2)\varphi(\xi)\, \rmd \xi
 \int_{-\infty}^{+\infty} \rme^{-\xi^2}\varphi(\xi)\, \rmd \xi - \\ \fl
 &\frac{1}{2\sqrt{\pi}} \int_{-\infty}^{+\infty} \rme^{-\xi^2}(1-2\xi^2)\varphi^2(\xi)\, \rmd \xi
 \bigg] ,
\end{eqnarray}
where the calculation was carried out up to the second perturbative order, $\Delta\lambda=\lambda_e - \lambda$, $\theta=2/(kw_0)$ is the $u_0$'s divergence, and $\xi=\sqrt{2}x/w_0$.

The simplest way to investigate the effect of the wavefront ripple is to consider the phase grating
\begin{equation}\label{varphi}
 \varphi(\xi)=\epsilon\sin(a\xi+\alpha) ,
\end{equation}
where $a=2\pi w_0/(\sqrt{2}\Lambda)$, $\Lambda$ is the grating pitch, and $\epsilon \ll 1$ rad. Hence, by carrying out the integrations in (\ref{corre-1}), we obtain \cite{Mathematica}
\begin{equation}\label{e1D}
 \frac{\Delta \lambda}{\lambda} \approx
 \frac{\theta^2}{8} \left[ 1 + \frac{a^2\rme^{-a^2}\cos(2\alpha) + (2+a^2)\rme^{-a^2/2}\sin^2(\alpha)}{2} \epsilon^2 \right]
\end{equation}

The $\theta^2/8$ term is proportional to the variance of the $u_0(x)$ angular spectrum. It is the one-dimensional equivalent of half the trace of the second central-moment of the angular spectrum \cite{Bergamin:1999,Mana:2017a}, which is the standard ingredient to calculate the needed correction and takes the diffraction of arbitrary paraxial beams into account.

\begin{figure}\centering
\includegraphics[width=6.3cm]{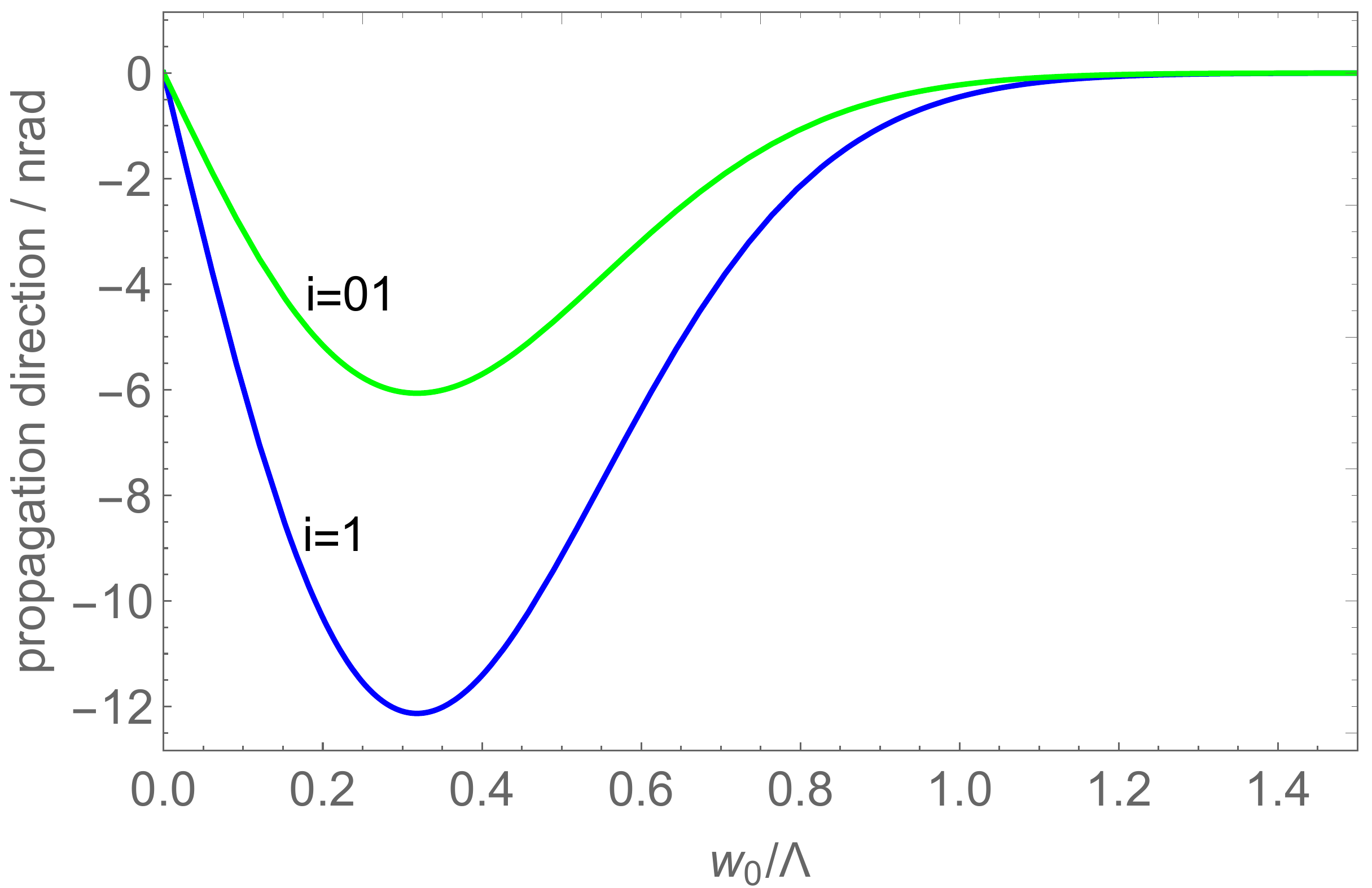}
\includegraphics[width=6.3cm]{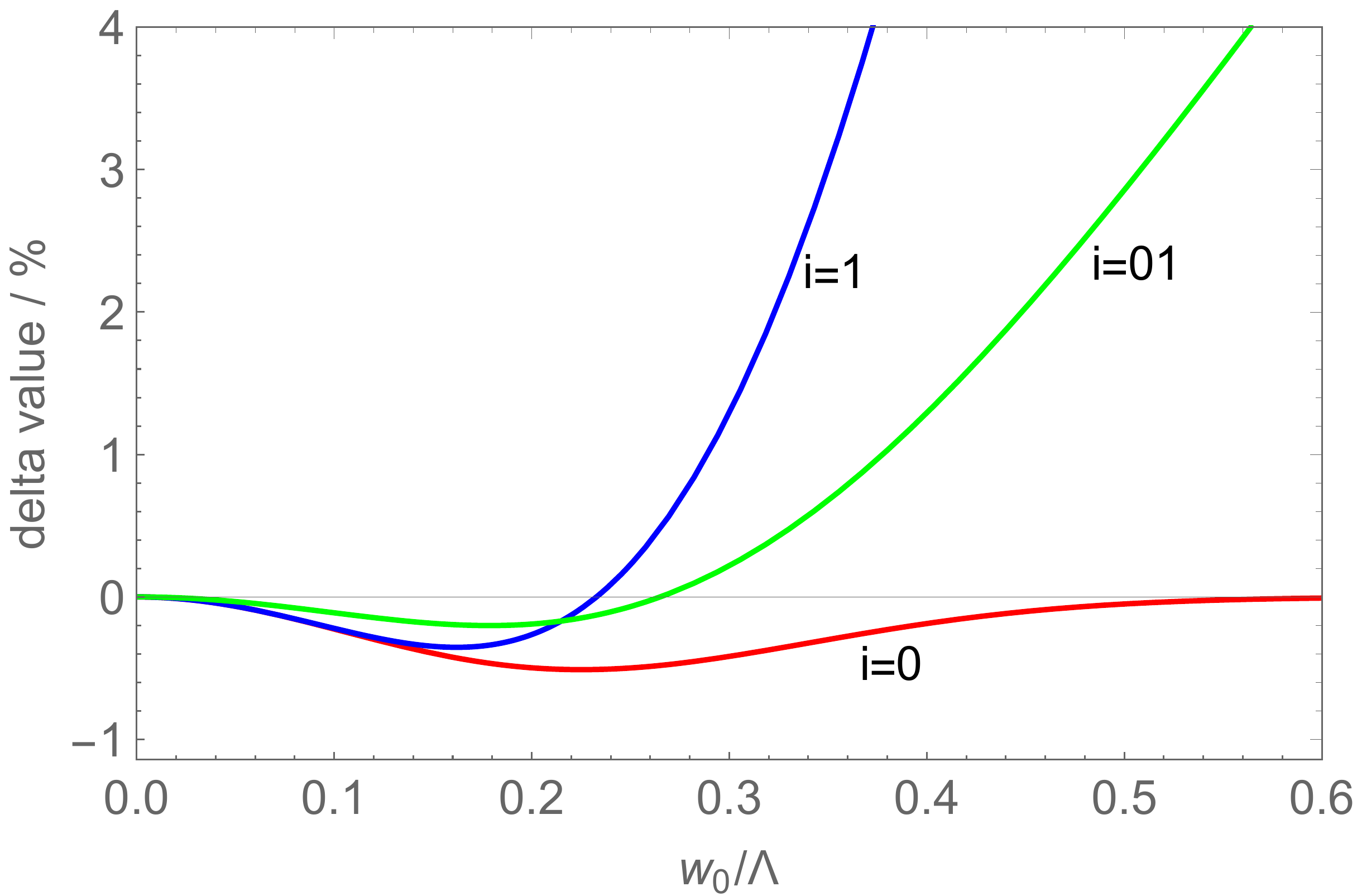}
\caption{Left: propagation directions of the aberrated and the superimposed beams ($u_1$ and $u_0+u_1$, respectively) exiting the interferometer. Right: delta values of the approximate wavelength differences (\ref{a1}-$c$) relative to the average difference (\ref{average}) {\it vs.} the fractional spatial frequency $w_0/\Lambda$ of the wavefront error (\ref{varphi}). The root-mean-square amplitude of the wavefront error is 10 nm.} \label{delta-1D}
\end{figure}

To quantify the impact of the wavefront error, we compare the fractional difference (\ref{e1D}) to the approximations $\Delta\lambda/\lambda \approx \Tr(\bGamma_i)/2$, where $\bGamma_i$ is: i) the second central-moment of the angular spectrum of the unperturbed beam $u_0$ illuminating the interferometer, ii) the aberrated, $u_1$, and iii) superimposed, $u_0+u_1$, beams exiting the interferometer. In the first case we have
\begin{numparts}\begin{equation}\label{a1}
 \Tr(\bGamma_0)/2 = \theta^2/8 ,
\end{equation}
in the second
\begin{equation}\label{a2}
 \Tr(\bGamma_1)/2 = \frac{\theta^2}{8} \left[ 1 + a^2(1+\rme^{-a^2} - 2\rme^{-a^2/2})\epsilon^2 \right] ,
\end{equation}
and, in the third,
\begin{equation}\label{a3}
 \Tr(\bGamma_{01})/2 = \frac{\theta^2}{8} \left[ 1 + \frac{1}{4}a^2(1+2\rme^{-a^2} - 2\rme^{-a^2/2})\epsilon^2 \right] .
\end{equation}\end{numparts}
In (\ref{a2}-$c$), we used the approximation (\ref{u-series}) and, for the sake of simplicity, set $\alpha=0$. It is worth noting that, as shown in Fig.\ \ref{delta-1D} (left), the propagation directions of the aberrated and superimposed beams, $u_1$ and $u_0+u_1$, deviate from that of the unperturbed beam $u_0$ by $\theta_1=-a\rme^{-a^2/4}\epsilon/k$ and $\theta_{01}=-a\rme^{-a^2/4}\epsilon/(2k)$, respectively. The misalignment occurring when $\Lambda/w_0 \approx 3$ mirrors the beam's perception of a wavefront tilt.

The fractional delta values of (\ref{a1}-$c$),
\begin{equation}\label{delta}
 \delta_i = \frac{\Tr(\Gamma_i)/2 - \Delta\lambda/\lambda} {\Delta\lambda/\lambda} ,
\end{equation}
relative to the fractional difference (\ref{e1D}) evaluated with $\alpha=0$,
\begin{equation}\label{average}
 \Delta\lambda/\lambda = \frac{\theta^2}{8} \left( 1 + \frac{a^2\rme^{-a^2} \epsilon^2}{2} \right) ,
\end{equation}
are shown in Fig.\ \ref{delta-1D} (right). In the case of a phase-grating pitch equal or shorter than the beam diameter, the increased angular spread of the aberrated beam does not affect the fringe period. We do not have an explanation of this.

\section{Numerical analysis}
The analytical treatment of section \ref{analytical} suggests that the actual fringe period might be different from the estimate based on the second central moment of the angular spectrum. Since the oversimplified analysis can hardly quantify this difference and the associated uncertainty, we resorted to a Monte Carlo estimate.

\begin{figure}\centering
\includegraphics[width=6.2cm]{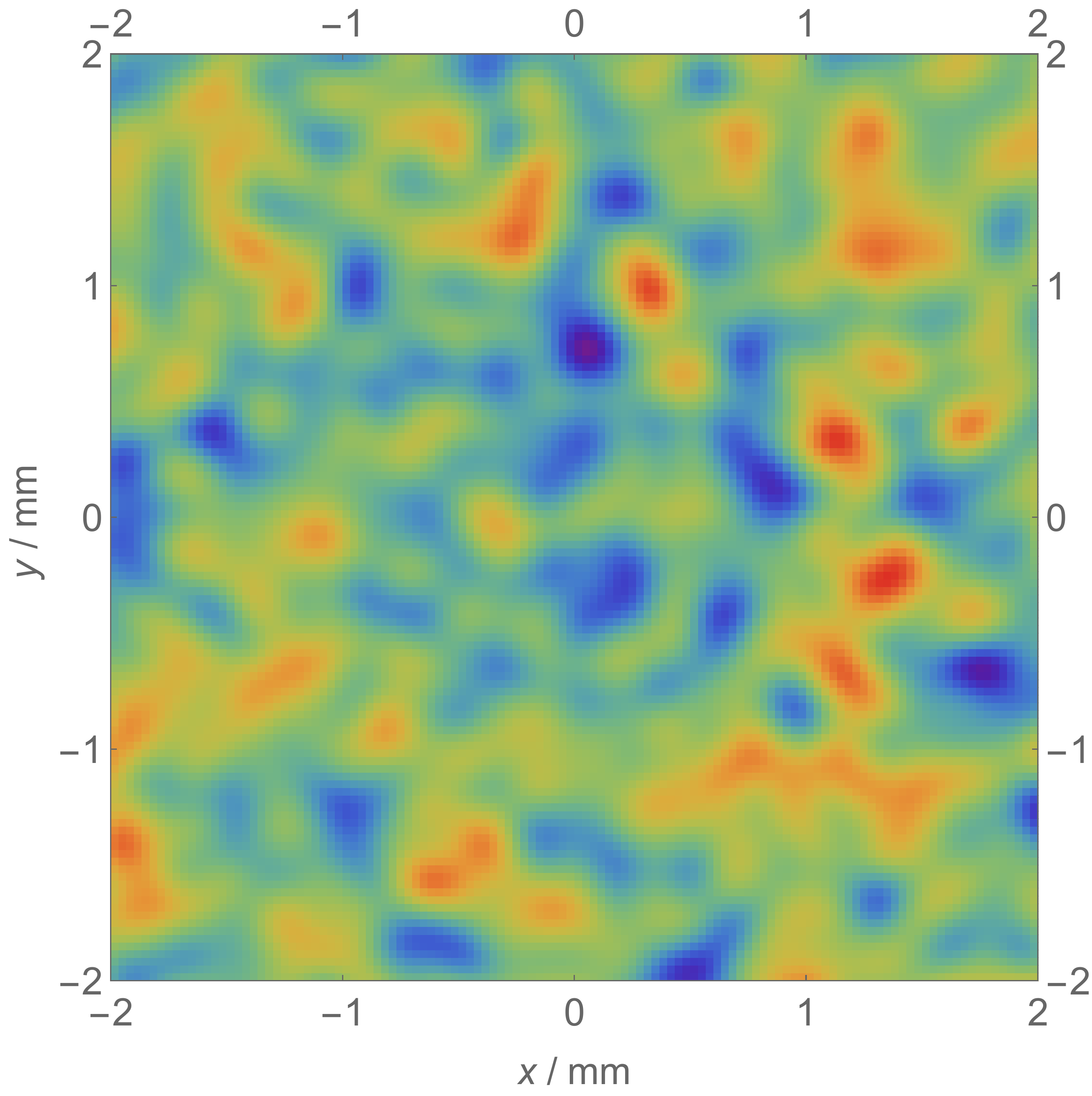}
\includegraphics[width=6.2cm]{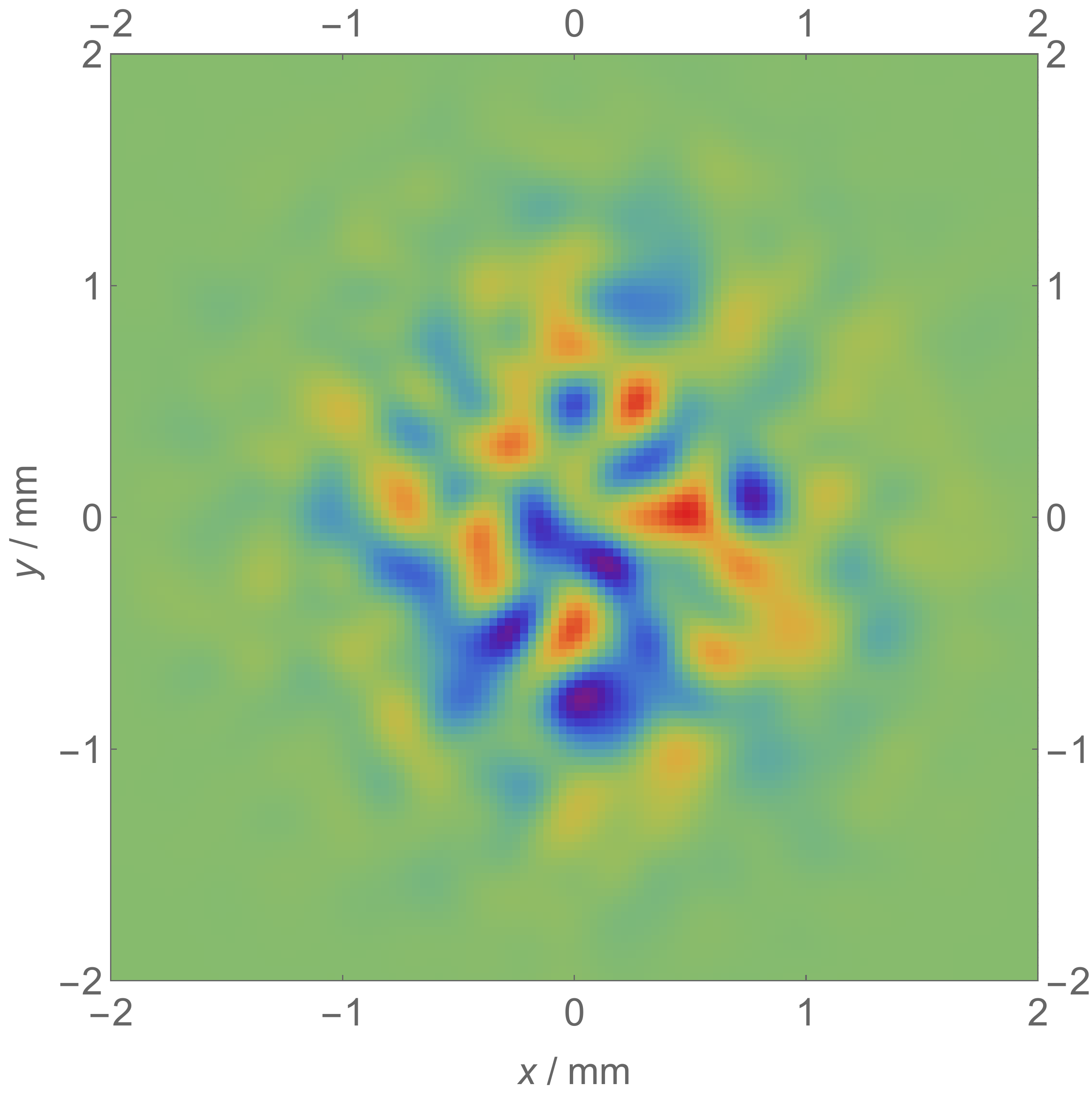}
\caption{Left: simulated wavefront error; the colour scale spans $\pm 30$ nm. Right: residuals from a Gaussian of the simulated intensity profile; the colour scale spans $\pm 2\%$ of the maximum beam intensity. The standard deviations of the wavefront errors and intensity profile are $\sigma_\varphi=10$ nm and $\sigma_A=0.025$, respectively.} \label{wfe-2D}
\end{figure}

\begin{figure}\centering
\includegraphics[width=8.5cm]{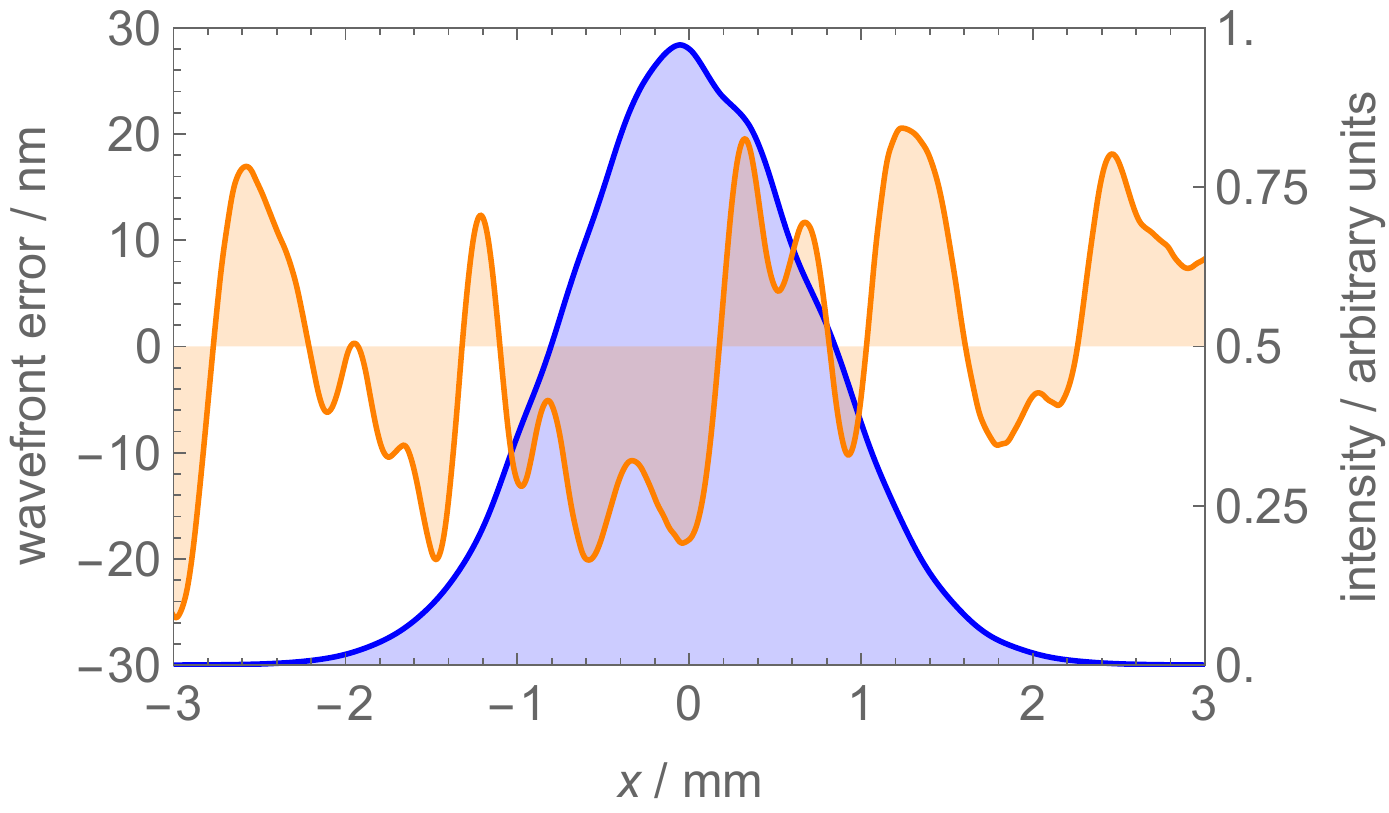}
\caption{Orange: $y=0$ section of the differential wavefront error shown in Fig.\ \ref{wfe-2D}. The blue line is the same section of the (aberrated) intensity profile.} \label{wfe}
\end{figure}

In the simulation, the two interfering beams,
\begin{equation}\label{MCu}
 u_i(x,y) = \big[ 1+A_i(x,y)/2 \big] g(x,y)\rme^{\rmi \varphi_i(x,y)} ,
\end{equation}
were independently generated $10^3$ times. In (\ref{MCu}), $A_i(x,y)$ and $\varphi_i(x,y)$ are the intensity and phase noises and
\begin{equation}
 g(x,y)=\rme^{-(x^2+y^2)/w_0^2}
\end{equation}
is the Gaussian beam feeding the interferometer, where $w_0=\sqrt{2}$ mm. {\colr In the Monte Carlo simulation, we considered collinear beams} and $A_i(x,y)$ and $\varphi_i(x,y)$ were collections of Gaussian, independent, and zero-mean random variables indexed by the observation-plane coordinates. As shown in Figs. \ref{wfe-2D} and \ref{wfe}, they have $\sigma_\varphi=10$ nm and $\sigma_A=0.025$ standard deviations and were filtered so has to have the same correlation length of about 0.5 mm observed experimentally \cite{Balsamo:2003,Sasso:2016}. {\colr We did not consider the wavefront curvature and imperfect recombinations of the interfering beams, which might be modelled by amplitude and phase perturbations \cite{Cavagnero:2006}.}

\begin{figure}\centering
\includegraphics[width=6.0cm]{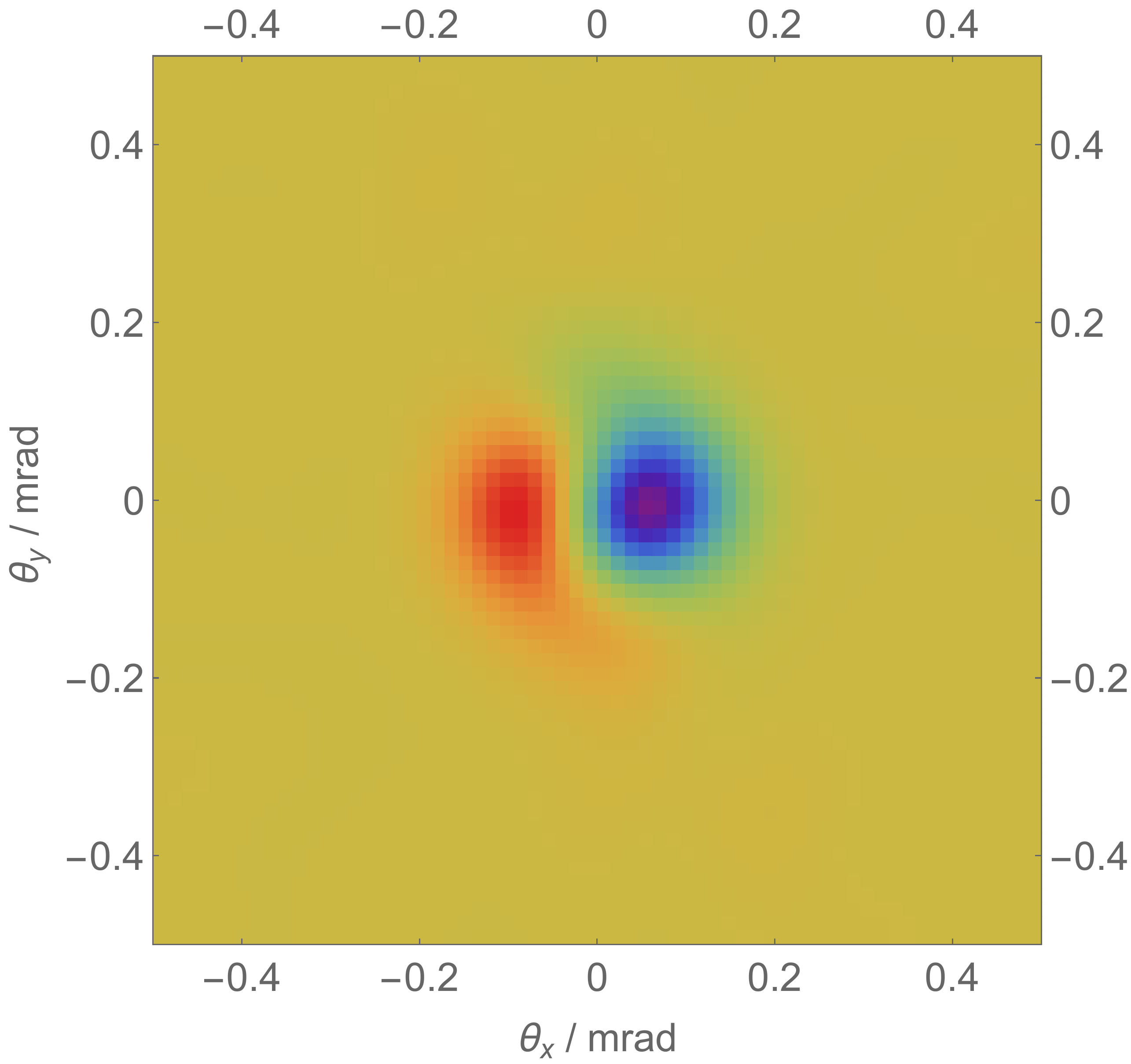}
\includegraphics[width=6.5cm]{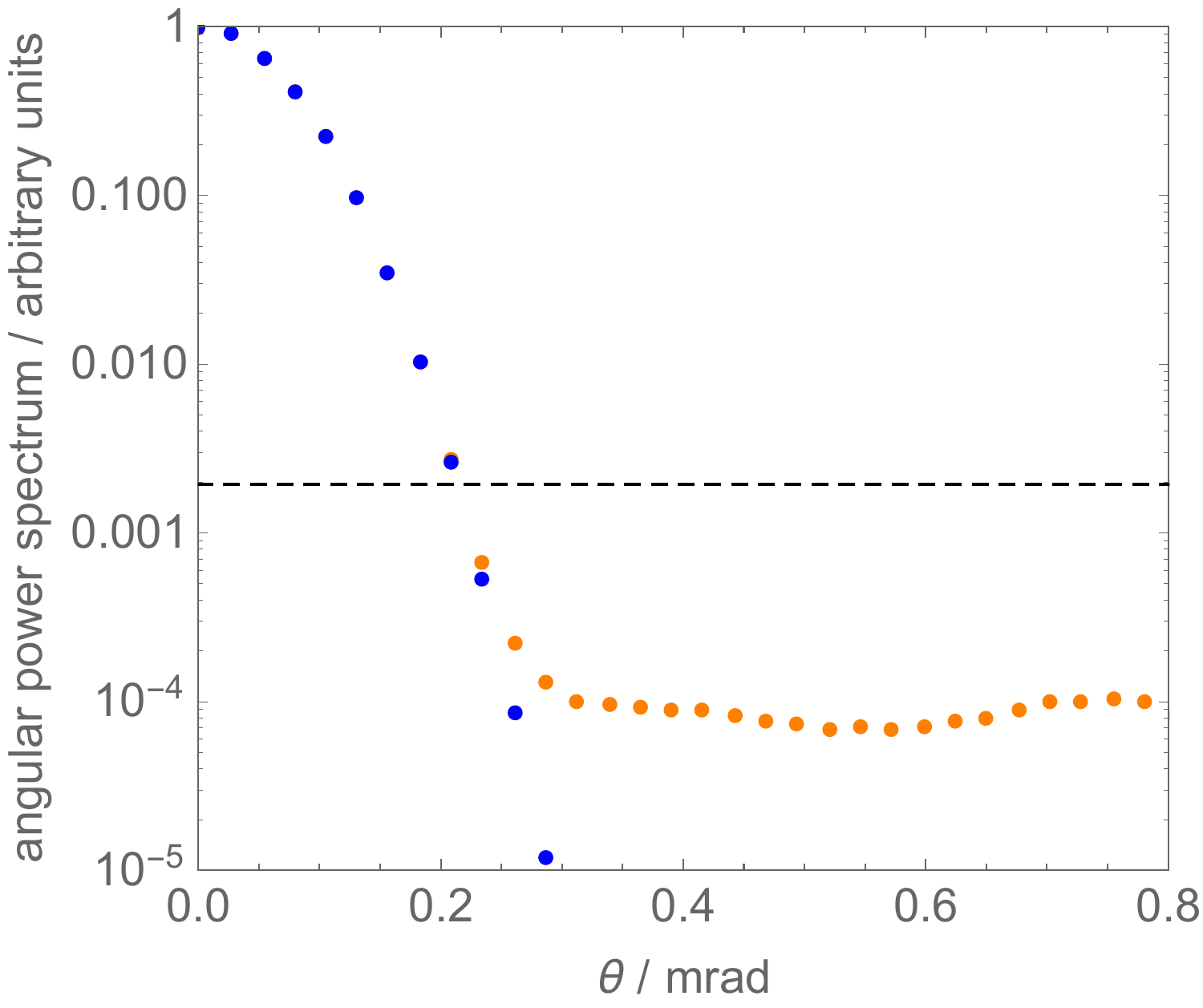}
\caption{Left: residuals of the angular power spectrum after subtracting the best-fit spectrum of a Gaussian beam; the colours indicate the normalized density. Right: averaged radial plot of the angular power spectrum (orange dots); the blue dots are the angular spectrum of the Gaussian beam feeding the interferometer. The standard deviations of the wavefront error and intensity profile are $\sigma_\varphi=10$ nm and $\sigma_A=0.025$, respectively. {\colr The dashed line indicates the instrumental background of the angular-spectrum measurements \cite{Mana:2017a}.}} \label{spectrum}
\end{figure}

The Monte Carlo simulation proceeded by Fourier transforming $u_0(x,y)$ and $u_1(x,y)$ and by calculating their interference and excess phase according to (\ref{Xi}) and (\ref{phase}). Figure \ref{spectrum} shows the angular spectrum of the aberrated beam. The plateau at $10^{-4}$ mrad$^{-2}$, which extends up to about $1$ mrad, originates from the $A(x,y)$ and $\varphi(x,y)$ noises. {\colr According to (\ref{error}), the Monte Carlo values} of the fractional difference between the fringe period and $\lambda$,
\begin{equation}\label{MC}
 \frac{\Delta\lambda}{\lambda}\bigg|_{\rm MC} = \frac{\Delta\Phi}{2\pi s/\lambda} ,
\end{equation}
where $\Delta\lambda=(\lambda_e - \lambda)/\lambda$, are obtained by propagating the fields back and forward by $s/2=\pm 50\lambda$ and by calculating the phase difference
\begin{equation}
 \Delta\Phi = \arg[\Xi(s/2)]-\arg[\Xi(-s/2)] ,
\end{equation}
which is null when the fringe period is equal to the plane-wave wavelength $\lambda$.

\begin{table}
\caption{\colr Comparison of analytical, equation (\ref{corre-1}), and numerical, equation (\ref{MC}), calculations of the fractional difference (expressed in nm/m) between the fringe period and the plane-wave wavelength in some one dimensional cases.}\label{test}
\begin{tabular}{@{}lllllll}
\br
 case &$\sigma_{A_1}$ &$\sigma_{A_2}$ &$\sigma_{\phi_1}$/nm &$\sigma_{\phi_2}$/nm &Eq.\ (\ref{corre-1}) &numerics \\
\mr
$A_1=A_0=0, \varphi_1=\varphi_0=0$                &$-$    &$-$    &$-$   &$-$    &1.792    &1.792 \\
$A_1=A_0 \ne 0, \varphi_1=\varphi_0 \ne 0$        &0.025  &0.025  &$50$  &$50$   &4.762    &4.762 \\
$A_0\ne 0, A_1=0, \varphi_0\ne 0, \varphi_1 =0$   &0.025  &$-$    &$10$  &$-$    &2.163    &2.152 \\
$A_0=0, A_1\ne 0, \varphi_0 =0, \varphi_1\ne 0$   &$-$    &0.025  &$-$   &$10$   &1.719    &1.724 \\
\br
\end{tabular}
\end{table}

To check the numerical calculations, we considered some one-dimensional cases, where the analytical expressions of the difference between the fringe period and the plane-wave wavelength are available, and compared the numerical calculations against the values predicted by (\ref{corre-1}). The results are summarized in table \ref{test}.

\begin{figure}\centering
\includegraphics[width=6.2cm]{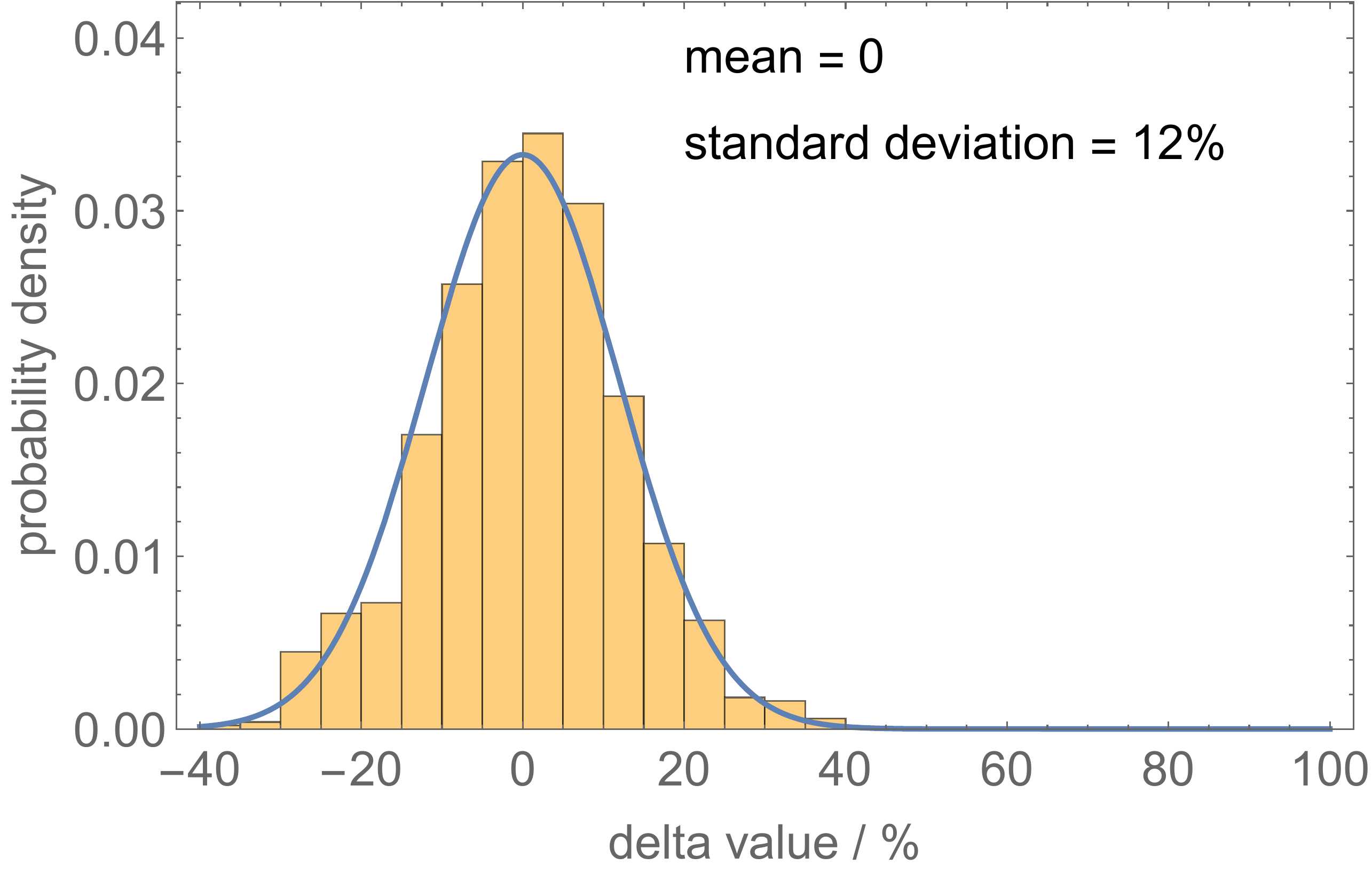}
\includegraphics[width=6.2cm]{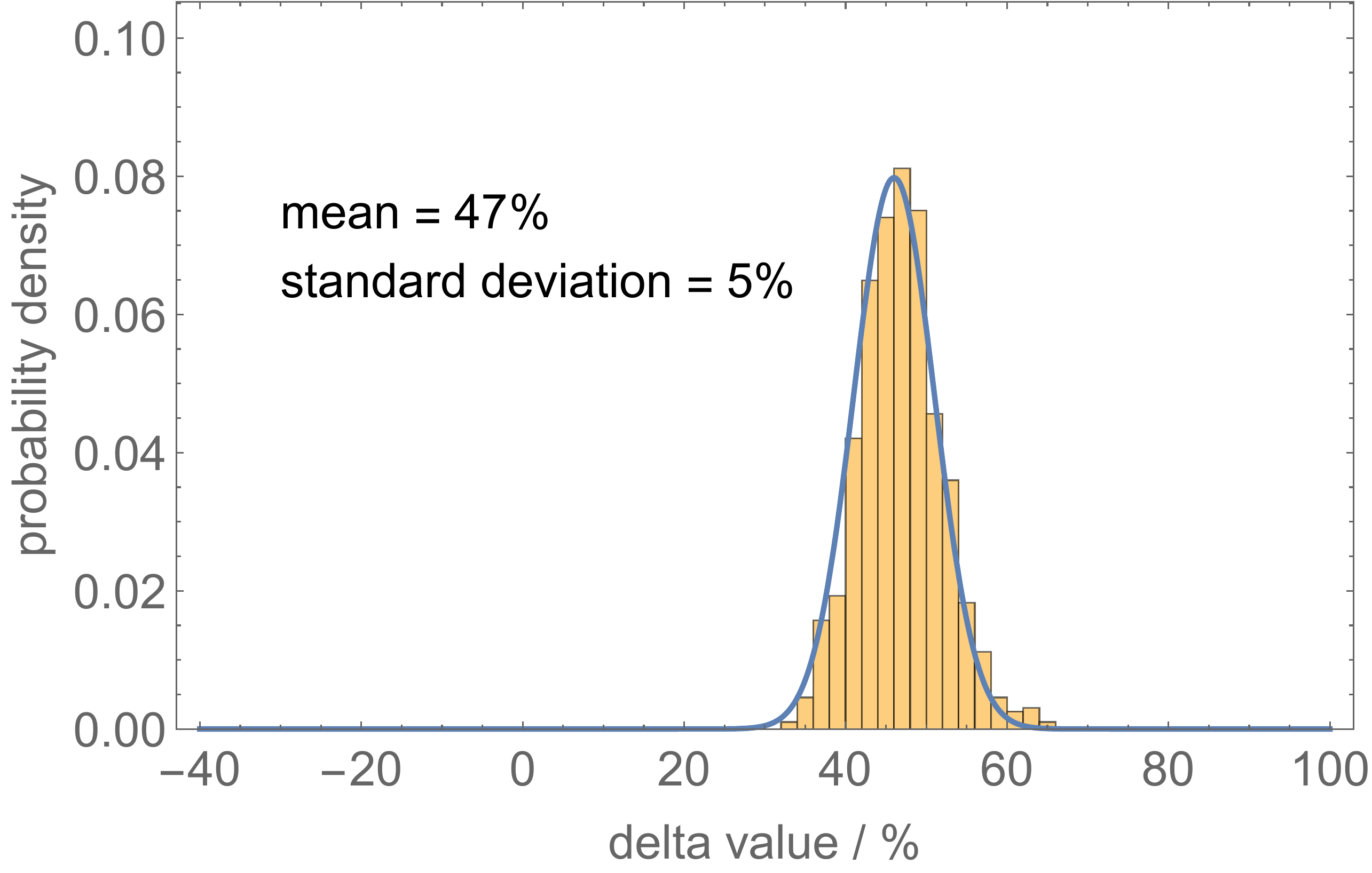}
\includegraphics[width=6.2cm]{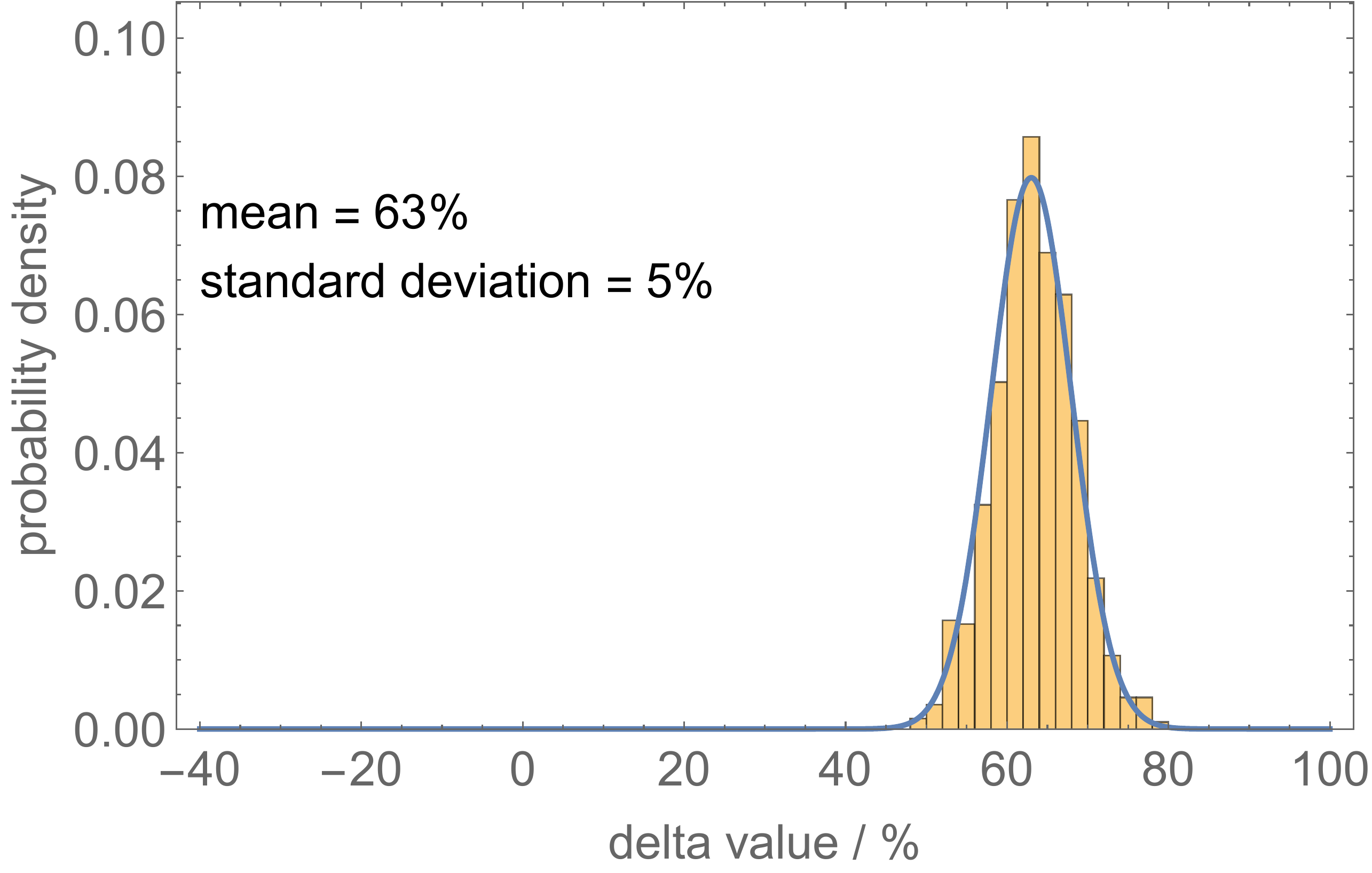}
\includegraphics[width=6.2cm]{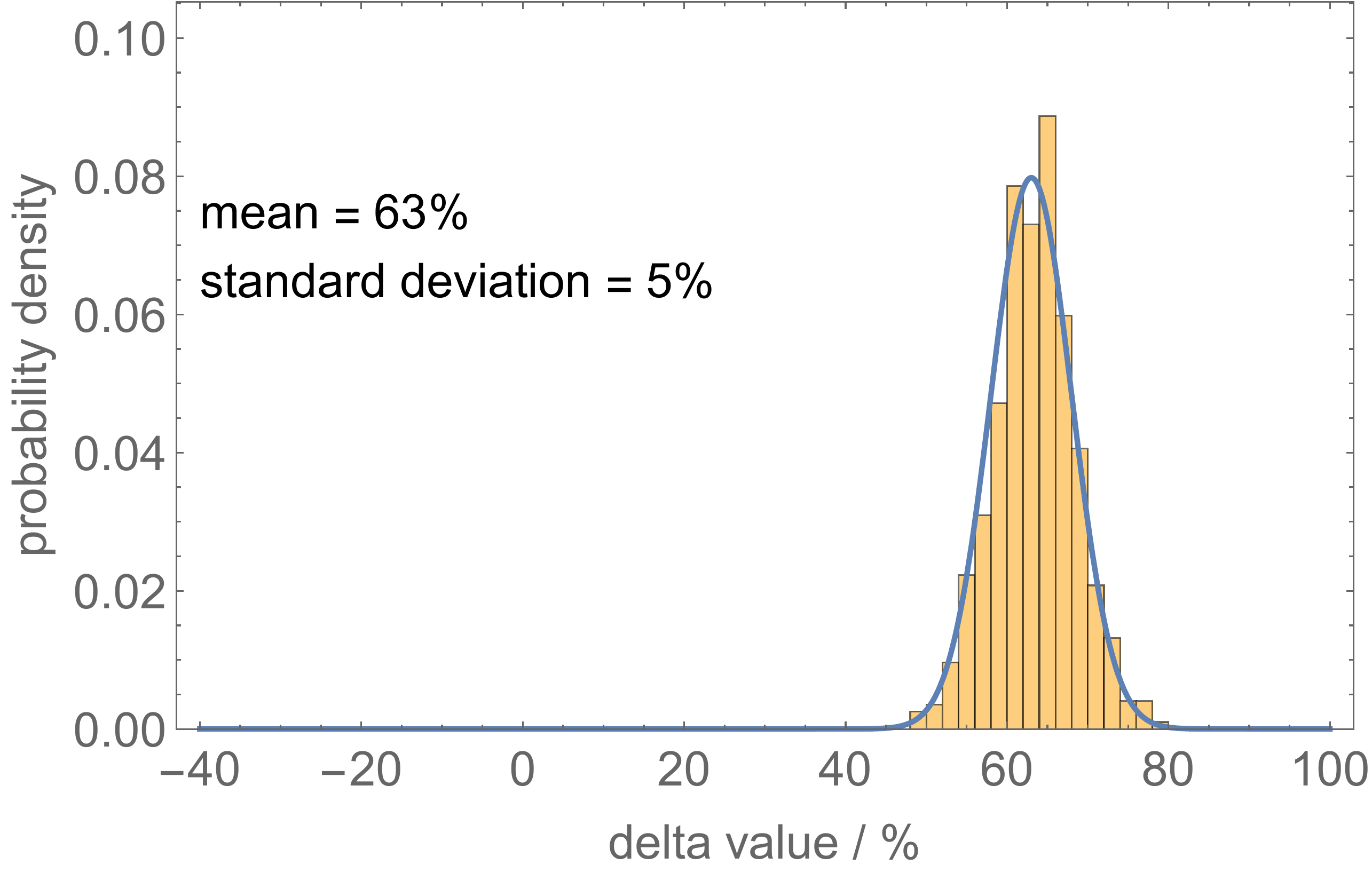}
\caption{Distributions of the delta values of the approximate wavelength differences (\ref{TrG}) -- obtained from the angular spectra of: the Gaussian beam feeding the interferometer, $|\tilde{g}(\bp)|^2$ (case 1, top left); the interfering-beam superposition, $|\tu_0(\bp) + \tu_1(\bp)|^2$, (case 3, top right); the beams leaving the interferometer, $|\tu_0(\bp)|^2$ and $|\tu_1(\bp)|^2$, (case 2, bottom left and right) -- relative to the numerical difference (\ref{MC}). The standard deviations of the wavefront errors and intensity profile are $\sigma_\varphi=10$ nm and $\sigma_A=0.025$. The blue line are the normal distributions best fitting the histograms.} \label{histo}
\end{figure}

To quantify the effect of two-dimensional wavefront errors, the fractional differences numerically calculated were compared to the approximations,
\begin{equation}\label{TrG}
 \frac{\Delta\lambda}{\lambda}\bigg|_i = \frac{1}{2}\Tr(\bGamma_i) ,
\end{equation}
which holds when the interferometer does not aberrate the interfering beams \cite{Bergamin:1999,Mana:2017a}, used to correct the interferometric measurements in \cite{Bartl:2017,Fujii:2018}. In (\ref{TrG}), $\bGamma_i$ is the second central-moment of the angular spectrum of: 1) the Gaussian beam feeding the interferometer, $|\tilde{g}(\bp)|^2$, 2) the interfering beams leaving the interferometer, $|\tu_0(\bp)|^2$ and $|\tu_1(\bp)|^2$, and 3) the interfering-beam superposition, $|\tu_0(\bp) + \tu_1(\bp)|^2$. The fractional delta values of the approximate differences (\ref{TrG}) relative to the numerical one (\ref{MC}),
\begin{equation}\label{chi}
 \delta_i = \frac{\Tr(\Gamma_i)/2 - \left(\Delta\lambda/\lambda \right)_{\rm MC}}{\left(\Delta\lambda/\lambda\right)_{\rm MC}} ,
\end{equation}
are shown in Fig.\ \ref{histo}.

The difference estimated from the angular spectrum of the beam entering the interferometer is equal to the mean of the actual (numerically calculated) values, which are scattered by about 12\%. Contrary, the differences estimated from the angular spectra of the beams leaving the interferometer, superposed or not, are significantly larger than truth. It is worth noting that the period values separately calculated from the angular spectra of each of the two beams leaving the interferometer are statistically identical. This agrees with the observation that the Monte Carlo simulation does not distinguish between the interfering beams.

\section{Conclusions}
We observed that the laser beams leaving the combined x-ray and optical interferometer used to measure the lattice parameter of silicon display wavelength and phase imprints having a spatial bandwidth of a few mm$^{-1}$ and local wavefront errors and wavelength variations as large as $\pm 20$ nm and $\pm 10^{-8}\lambda$ \cite{Sasso:2016}. These aberrations are likely due to the interferometer optics. Besides, the observed imprints correspond to a root-mean-square deviation from flatness of each of the optics' surfaces of less than 3 nm for scale lengths from 0.1 mm to 2 mm. {\colr

Since our measurements, which were corrected on the basis of the angular spectra of the laser beam, aimed at $10^{-9}$ fractional accuracy, questions arise about the impact of these errors. The Monte Carlo simulation of the interferometer operation indicates that the corrections made depend on the angular spectra having been measured before or after the interferometer. The correction is faithfully evaluated when the spectrum is measured before the interferometer.

Unfortunately, not being aware of the problem, we measured the angular spectra after the interferometer \cite{Massa:2011,Massa:2015,Mana:2017a}. However, we note that the excess of correction is due to the angular-spectrum plateau that increases the central second-moment of the incoming beam. Since, as shown in Fig.\ \ref{spectrum} (right), this plateau is indistinguishable from the instrumental background of the spectrum measurement, it was subtracted from the data and excluded from consideration. Therefore, the plateau and, consequently, the wavefront errors did not bias the corrections made. By using the typical aberrations observed in our set-up, the correction uncertainty is 12\%, which is within the 15\% cautiously associated with them \cite{Mana:2017a}.

Nevertheless this reassuring conclusion, our work evidenced unexpected critical issues, which deserve further investigations and on-line determinations of the needed correction, e.g., by reconstructing its value from the (measurable) dependence on the detector area.} These results have a value also in other experiments, such as those determining the Planck constant and the local acceleration due to gravity, where precision length-measurements by optical interferometry play a critical role.


\section*{References}
\providecommand{\newblock}{}

\end{document}